\documentstyle[12pt]{article}
\begin{document}
\baselineskip 28pt

\begin{center}
{\bf RADIATIVE DECAYS OF HEAVY MESONS AND THE DETERMINATION OF THE
STRONG g-COUPLING}\footnote{Presented at the XXXIX Cracow School
of Theoretical Physics, May 29-June 8, 1999, Zakopane, Poland}\\ \ \\
PAUL SINGER\\
Department of Physics, Technion-Israel Institute of Technology, Haifa, Israel
\end{center}

The strong $g$-coupling characterizes the interaction of heavy mesons with
pions in typical vertices $H^*H\pi$, $H^*H^*\pi$, where $(H^*;H)$ stands for
vector and pseudoscalar $(B^*;B)$ or $(D^*;D)$ heavy mesons. Its 
estimation by different theoretical methods has led to a wide range of
possible values. We describe a new approach to the determination of $g$, which
exploits the rare radiative decays $B^*\rightarrow B\gamma\gamma$ and 
$D^*\rightarrow D\gamma\gamma$. It is shown that the branching ratio
of $D^*\rightarrow D\gamma\gamma$ can be expressed as a function of a single
unknown $g$ and we calculate it to be in the measurable range between
$1.6\times 10^{-6}$ and $3.3\times 10^{-5}$ for $0.25 < g < 1$.

\pagebreak
\begin{center}
{\bf 1. Introduction and Experimental Overview}
\end{center}

We consider here the strong and electromagnetic decays of the heavy
vector mesons $D^*$ and $B^*$ of spin-parity $1^-$, with the special aim
of gaining information on their strong couplings $g_{B^*B\pi}$, $g_{D^*D\pi}$.
It is well known that the main decays of $D^*$'s proceed either as a strong
transition $D^*\rightarrow D\pi$ with a final pion of about 40
MeV/c momentum, or as an electromagnetic transition $D^*\rightarrow
D\gamma$ with
a photon of momentum of nearly 140 MeV/c. On the other hand, since the mass
difference $M_{B^*}-M_B$ is only 45.8 MeV, the decay
$B^*\rightarrow B\pi$ cannot take place and the main decay of $B^*$ is the
radiative process $B^* \rightarrow B\gamma$.

The interaction between mesons containing a single heavy quark $Q$ and
pseudoscalar 
Goldstone bosons
is presently best described by an effective theory [1,2,3] which contains
flavour and spin symmetries in the heavy
mesons sector and chiral SU(3)$_L \otimes SU(3)_R$ symmetry in the light one
(for a recent review, see [4]). As a result of the heavy quark symmetry,
the heavy meson chiral Lagrangian ($HM\chi L$) which implements this scheme
contains
just one coupling, denoted by $g$, to characterize the strength of 
several vertices, $D^*D^*\pi$, $D^*D\pi$, $B^*B^*\pi$ and $B^*B\pi$. While the
$g_{B^*B\pi}$ vertex is not accessible in a direct decay, the $g_{D^*D\pi}$
coupling
may be measured in principle from the decay width of the decay channel
$D^*\rightarrow D\pi$. So
far, only an upper limit is available from the ACCMOR Collaboration at CERN
[5], 
$\Gamma(D^{*+})$ $<131$ KeV based on the high-resolution measurement
of 127 $D^{*+}$ events. 

In view of the great interest in the strength of the $B^*B\pi$ and
$D^*D\pi$ vertices, which are
relevant for the analysis of various $B$ and $D$ decays, a large number of
calculations
has been performed to obtain the $g$-coupling. The results are quite
divergent,
as it will be described in the next chapter. Recently, Daphne Guetta and
myself [6] have suggested
a new approach to the determination of $g$. We considered the decays
$B^*\rightarrow B\gamma\gamma$ and $D^*\rightarrow D\gamma\gamma$ within the
framework of $HM\chi L$ and we have shown that the measurement of the 
branching ratios and spectra of these decays can provide information on
the $g$-coupling. This method is especially profitable in the charm
sector.

As a background for the model we have built, I review here succinctly the
experimental status of the $B^*$ and $D^*$ decays and present a
summary of the theoretical attempts to describe the main decays to 
a pion or a photon.

The main decay of $B^*$, the electromagnetic transition $B^*\rightarrow
B\gamma$
has been observed both at the Cornell Electron Storage Ring (CESR) [7]
and at LEP [8]. There is obviously no measurement
of the width of this transition. It has been studied in a variety of 
theoretical models including quark models [9], the chiral bag model [10]
followed
by effective chiral Lagrangian approaches [11], potential models [12], QCD
sum rules 
[13] and an analysis of experimental $D^*$ branching ratios using $HM\chi L$
[14]. The predictions range from $\Gamma(B^{*o}(B^{*+})\rightarrow 
B^0(B^+)\gamma) = 0.04(0.10)$ KeV [13] to 0.28(0.84) KeV [10,11], with
most calculations concentrating in the higher range.

The $D^*$ meson was discovered more than twenty years ago [15] and its
decay branching ratios have been studied extensively. The current PDG
averages [16] are Br$(D^{*+}\rightarrow D^+\pi^o):{\rm Br}(D^{*+}\rightarrow
D^o\pi^+):{\rm Br}(D^{*+}\rightarrow D^{+}\gamma) = 
(30.6\pm2.5)\%:(68.3\pm1.4)\%:(1.1\pm 2.1 \pm 0.7)\%$ and
$Br(D^{*o}\rightarrow D^o\pi^o):
{\rm Br}(D^{*o}\rightarrow D^o\gamma)= (61.9\pm 2.9)\%:(38.1\pm2.9)\%$.
A recent [17] CLEO experiment on $D^{*+}$ decays gives the more accurate
branching ratios Br$(D^{*+}\rightarrow D^+\pi^o):{\rm Br}
(D^{*+}\rightarrow D^o\pi^+):{\rm Br}(D^{*+}\rightarrow D^+\gamma)=
(30.7\pm0.7)\%:(67.6\pm0.9)\%:(1.7\pm0.6)\%$.

These $D^*$ electromagnetic and strong decays have been calculated with the
same models used for $B^*$ and most papers of Refs. [9]-[14] have
considered also $D^*$ decays. Additional calculations referring to 
$D^*$ only include quark models [18], the chiral bag model
[19] and use of sum rules [20]. Here again the theoretical 
calculations span an order of magnitude range for the prediction of the
widths,
from $\Gamma(D^{*o})\simeq (3-10)$KeV [13] to (60-120)KeV [9,11]. Many of
these theoretical results are fitted to obtain correct relative widths.
The real test will come when the absolute width will be measured. According
to several of the models, typical values are $\Gamma(D^{*+} \rightarrow \
{\rm all}) \simeq 80$KeV, $\Gamma(D^{*o}\rightarrow \ {\rm all})\simeq 60$
KeV [9,18,19], not far from the present upper limit [5].\\

\begin{center}
{\bf 2. The Heavy Meson Chiral Lagrangian and Estimation of {\it g} }
\end{center}

The treatment of the physical processes involving soft pions we described
in the previous section is performed within the framework of an effective
theory, the ``Heavy Meson Chiral Lagrangian'' ($HM\chi L$) which embodies
two principal symmetries of Quantum Chromodynamics. At one end, there is
the SU(2N$_f$) heavy flavour-spin symmetry characteristic of the infinite
heavy quark mass limit. In this limit, the interactions with the
light hadrons are independent of the mass of the heavy quarks; moreover,
the independence
of the interaction on the spin $s_Q$ of the heavy quark allows to define
degenerate doublets of heavy states with spin-parity $s^P_Q=(s_\ell\pm
\frac{1}{2})^P$ where
$s_\ell$ is the spin of the light quark, which in our case is $q=u,d,s$.

The doublet members are the pseudoscalar and vector mesons corresponding
to $s_\ell=\frac{1}{2}$ and we make the assumption that both the $c$ and
the $b$
quark are sufficiently heavy. Symmetry-breaking corrections to this
scheme are taken into account by terms obtained in a $\frac{1}{M_Q}$
expansion [4].

At the other end of the energy scale is the chiral limit of QCD realized
for $M_\ell\rightarrow 0(\ell=u,d,s)$; the QCD Lagrangian is then invariant
under
SU(3)$_L \times {\rm SU}(3)_R$ transformations. This symmetry is 
spontaneously broken to the vector subgroup SU(3)$_V$ and the resulting 
Goldstone bosons are the eight pseudoscalar mesons $\pi, K$ and $\eta$.
The quark mass terms breaking the chiral symmetry help the Goldstone bosons
to 
acquire their mass.

The effective Lagrangian which contains these symmetries is expressed [1-4]
in terms of the hadronic fields for heavy and light mesons. The heavy vector
($B^*,D^*)$ and pseudoscalar $(B,D)$ mesons are represented by a $4\times 4$
Dirac matrix $H$, with one spinor index for the heavy quark and the second
one 
for the light degree of freedom,
\begin{eqnarray}
H=\frac{1+ v\hspace{-0.20cm}/ }{2} \left[P^*_\mu v^\mu - P\gamma_5\right],
\ \ \
\bar{H} = \gamma_o H^\dagger\gamma_o \ . %1 
\end{eqnarray}
$P^*_\mu$ and $P$ are the respective annihilation operators of vector
($1^-$) and
pseudoscalar $(0^-$) heavy mesons with four-velocity $v_\mu$.

The Goldstone bosons are represented with the aid of a unitary $3 \times 3$
matrix
$\sum=\exp(2i M/f)$ with the $M$ being the usual $3\times 3$ hermitian
traceless matrix
describing the octet of pseudoscalar bosons.

The most general Lagrangian describing the interaction of heavy mesons with
Nambu-Goldstone
bosons, which is invariant under Lorentz transformations, parity, heavy-quark 
spin flavour symmetry
and chiral symmetry is given in the framework we described by [1-4]
\begin{eqnarray}
{\cal L} = && i Tr\left\{ \bar{H}_a v^\mu D_{\mu ba} H_b \right\} +
\frac{f^2}{8} \partial^\mu \Sigma_{ab} \partial_\mu \Sigma^\dagger_{ba}
\nonumber \\
&& ig Tr\left\{\bar{H}_a\gamma_\mu \gamma_5 A^\mu_{ab} H_b\right\}  %2
\end{eqnarray}
where $D_\mu$ is a covariant derivative. $D_\mu = \partial_\mu + V_\mu$, 
and $f$ is the pion decay constant, $f=132$ MeV.

Using $\Sigma(x) = \xi^2(x)$ one introduces a vector $V_\mu =\frac{1}{2}
(\xi^\dagger\partial_\mu\xi - \xi\partial_\mu\xi^\dagger)$ and an axial
current
$A_\mu =\frac{1}{2}
(\xi^\dagger\partial_\mu\xi + \xi\partial_\mu\xi^\dagger)$. The first two
terms of (2) are the
kinetic energy terms and the third one is an interaction term, defining the
strong-interaction
coupling $g$. Explicit interaction terms are obtained by expanding the axial
current in (2) and keeping the first term $A^\mu = (i/f)\partial_\mu M +
\dots$.
Expressing the interaction in terms of $(D^*,D)$ fields (the same holds for
$B^*,B$ fields) one has
\begin{eqnarray}
{\cal L}^{\rm int}_{\rm eff} =
\left[ - \frac{2g}{f} D^*_\mu \partial^\mu M D^\dagger + h.c\right] + 
\frac{2gi}{f} \epsilon_{\mu\nu\sigma\tau} D^{*\mu} \partial^\sigma M
D^{*\dagger\nu} v^\tau %3
\end{eqnarray}
and we see that, e.g., $D^*D\pi$ and $D^*D^*\pi$ vertices are characterized
with the
same strength $g$. The $g$-coupling is directly related to the hadronic
coupling 
$g_{D^*D\pi}$ which is defined by the on-shell matrix element
\begin{eqnarray}
\langle {D^o} (p) \pi^+(q)|D^{*+}(p+q)\rangle = g_{D^*D\pi}\epsilon_\mu
q^\mu %4
\end{eqnarray}
where $\epsilon^\mu$ is the polarization vector of $D^{*+}$. Likewise,
\begin{eqnarray}
\langle {D^{*o}} (p, \epsilon_1)  \pi^+(q)|D^{*+}(p+q, \epsilon_2)\rangle = 
g_{D^*D^*\pi}\epsilon_{\mu \nu\sigma \tau} \epsilon_1^\mu\epsilon^\nu_2
p^\sigma q^\tau \ .
%5
\end{eqnarray}
>From (3-5) one finds the link
\begin{eqnarray}
g_{D^*D\pi} = g_{D^*D^*\pi} = \frac{2M_D}{f} g  \ . %6
\end{eqnarray}
Moreover, isospin symmetry requires
\begin{eqnarray}
g_{D^*D\pi} \equiv g_{D^{*+}D^o\pi^+} = 
- \sqrt{2}g_{D^{*+}D^+\pi^o} = 
\sqrt{2}g_{D^{*o}D^o\pi^o} = 
- g_{D^{*o}D^+\pi^-} \ . %7
\end{eqnarray}

The interest in the value of $g$ is not limited to the theoretical interest
in the 
strength of the axial interaction defined in Eq.~(2). The knowledge of $g$
is of great
phenomenological value, since its strength is required in the analyses of
many electroweak
processes [4]. Among these, for example, are heavy-to-light semileptonic
exclusive decays
like $B\rightarrow \pi \ell \nu$, $D_s \rightarrow K \ell \nu$ which are
promising 
processes for the extraction of $CKM$ matrix elements like $|V_{ub}|$, and
their analysis with
VMD requires the knowledge of $g_{B^*B\pi}$. Other processes requiring such
knowledge are
$B\rightarrow D^*\pi\ell\nu$ decays, chiral corrections to $B\rightarrow D$
processes,
decay constants of heavy mesons,
radiative processes like $D^*\rightarrow D\gamma$, $B^*\rightarrow B\gamma$
and
more. Thus, no wonder that a large number of theoretical papers has been
devoted during the last
years to the computation of $g$. In the rest of this Section we present a
succint overview of the
main attempts in this direction.

The theoretical attempts may be grouped into several classes; however, even
within a single 
class of models,
the variation of $g$ turns out to be quite large. The non-relativistic
quark model leads to 
[3] the largest value $g=1$, while slightly modified quark models [18, 21]
bring this value
down to about $g=0.8$. On the other hand, in a calculation [22] in which
the effect of the
relativistic motion of the light antiquark is taken into account by the use
of Salpeter equation
one arrives at $g=\frac{1}{3}$. Somewhat higher values were obtained in
recent quark-model 
calculations:
with a relativistic quark model based on the light front formalism Jaus
obtains [9] $g=0.56$;
in a relativisitic quark model with direct quark-meson interactions [23]
one arrives at $g=0.46$;
and a quark model with Dirac equation [24] finds $g=0.61$.  The QCD sum
rules have also been
used [25] extensively to the calculation of $g$. The outcome of these
calculations is
generally in the direction of small $g$ values, between 0.15 and 0.35,
which would imply
that the decay width of $D^*$ is rather small, i.e. below 45 KeV. Finally, we
mention a recent lattice QCD determination [26] of $g=042(4)(8)$ and the
analysis of 
Stewart [14] of the experimental data on $D^*\rightarrow D \pi, D\gamma$
which incorporates
symmetry breaking terms in the Lagrangian and deduces
$g=0.27^{+0.09}_{-0.04}$.\\

\begin{center}
{\bf 3. A Model for Two-photon Decays $D^*\rightarrow D\gamma\gamma$,
$B^*\rightarrow B\gamma\gamma$}
\end{center}

Recently we have considered the two-photon processes $B^*(D^*)\rightarrow
B(D)\gamma\gamma$ [6], not
discussed previously in the literature, by using the $HM\chi L$ and we
found that the
$D^*\rightarrow D\gamma\gamma$ could provide a measurement of the much
sought after $g$-coupling.

The calculation of radiative processes requires the addition of the
electomagnetic interaction
to the Lagrangian of Eq.~(2). This is performed by the usual procedure of
minimal coupling 
which leads to the replacement of derivative operators
by covariant derivatives containing the photon field. However, this does
not suffice to 
account for the observed magnetic dipole transitions 
$B^*\rightarrow B\gamma$, $D^*\rightarrow D\gamma$; to 
account for these, a contact gauge invariant electromagnetic term
proportional to
$F_{\mu\nu}$ must be added, which has the form in the heavy mass limit [11,
14]
\begin{eqnarray}
{\cal L}^{(\mu)} = \frac{e\mu}{4} Tr (\bar{H}_a \sigma_{\mu\nu}F^{\mu\nu}
H_b \delta_{ab}), %8
\end{eqnarray}
where $\mu$ is the strength of this anomalous magnetic dipole interaction
and has mass
dimension $[1/M]$.

In the following, we present our calculation for the charm sector; then, we
shall comment on
the features arising in the beauty sector. From (8), additional
electromagnetic vertices
obtain representing $D^*D^*\gamma$ and $D^*D\gamma$ interactions of same
strength. These are
\begin{eqnarray}
\langle \gamma (k,\epsilon){D}(v_1)|D^*(v_2,\epsilon_2)\rangle =
-ie M_{D^*}\mu\epsilon_{\mu\nu\alpha\beta}\epsilon^\mu k^\nu v_2^\alpha
\epsilon^\beta_2 %9
\end{eqnarray}
\begin{eqnarray}
\langle\gamma(k,\epsilon){D^*}(v_1,\epsilon_1)|D^*(v_2,\epsilon_2)\rangle =
e \mu M_{D^*}(\epsilon_1\cdot k \epsilon \cdot \epsilon_2 - \epsilon_2\cdot
k\epsilon \cdot \epsilon_1)
 \ . %10
\end{eqnarray}\
The calculation is performed to leading order in chiral perturbation theory
and to this order 
there are
no counterterms [14,27]. In addition to the Feynman diagrams obtained from
Eq.~(2) 
with minimal electromagnetic
interaction and from Eq.~(8), we must include to the same order the pion
axial anomaly,
whose strength is known. All these terms are of the same order in an
$1/N_c$ expansion.

We start with the description of the decay of the neutral $D^*$. The decay
amplitude $A$ for
$D^{*o}\rightarrow D^o\gamma\gamma$ may be written as
\begin{eqnarray}
A = A_{\rm anomaly} + A_{\rm tree} + \sum^6_{i=1} A^{(i)}_{\rm loops} \ . %11
\end{eqnarray}
We shall describe now the eight contributions to the amplitude, indicating
the couplings
entering into each of them, without giving here the detailed expressions
which can be found
in Ref. [6].

$A_{\rm anomaly}$ represents $D^{*o}\rightarrow D^{o\prime\prime}
\pi^{\prime\prime}
\rightarrow D^o\gamma\gamma$ via a virtual neutral pion. 
Since the physical decay $D^{*o}\rightarrow
D^o\pi^o$ is allowed, we limit ourselves to a region for $s=(k_1 + k_2)^2$
which goes 
up to 20 MeV away from  the pion mass. Given the strength of the pion axial
anomaly
of $(\alpha/\pi f)$, where $f$ is the pion decay constant and
$\alpha=e^2/4\pi$, 
$A_{\rm anomaly}$ is proportional to $\alpha g_{D^*D\pi}$. The tree level
graph is
due to the transition $D^{*o}\rightarrow \ ^{\prime\prime}D^{*o}~
^{\prime\prime} \gamma 
\rightarrow D^o\gamma\gamma$, containing two insertions of the anomalous
magnetic operator.
Hence, $A_{\rm tree}$ is proportional to $\alpha\mu^2$. $A^{(1)}_{\rm
loop}$ describes
the transition $D^{*o}\rightarrow(D^{*+}\pi^-)\rightarrow D^o\gamma\gamma$
with the two photons
radiated from the virtual charged pion. Additional graphs, required by
gauge invariance have
one photon radiated from the loop and the second emitted from the
$D^*D^*\pi$, $D^* D\pi$ 
vertices 
or both photons emitted from these vertices. The other loop diagrams,
$A^{(2)}_{\rm loop} -
A^{(6)}_{\rm loop}$ come from diagrams where both the strong coupling and
the magnetic one
are involved. $A^{(2)}_{\rm loop}$ is given by $D^{*o}\rightarrow "D^{*o}"
\gamma \rightarrow
(D^{*+}\pi^-)\gamma\rightarrow D^o\gamma\gamma$,
being thus proportional to $\alpha g_{D^*D^*\pi}g_{D^*D\pi}\mu$.
In $A^{(3)}_{\rm loop}$ a $D^{*o}D^o\gamma$ vertex replaces the
$D^{*o}D^{*o}\gamma$ one in the 
initial step, the diagram being proportional to $\alpha g^2_{D^*D\pi}\mu$.
$A^{(4)}_{\rm loop}$
describes the transition $D^{*o}\rightarrow(D^+\pi^-)\rightarrow
D^{*o}\gamma \rightarrow
D^o\gamma\gamma$, where the first photon is emitted by the charged pion from 
the $(D^+\pi^-)$ loop,
while the second one comes from the 
$D^{*o}\rightarrow D^o\gamma$ 
transition. The expression is then proportional to $\alpha
g^2_{D^*D\pi}\mu$. Exchanging
$D$ with a $D^*$ in the loop one gets $A^{(5)}_{\rm loop}$, which is thus
proportional to
$\alpha g^2_{D^*D^*\pi}\mu$. Finally, we have a diagram given by
$D^{*o}\rightarrow (D^+\pi^-)$
and then the virtual pion emits one photon while the virtual $D^+$ emits
the other one becoming
$D^{*+}$. The virtual $(D^{*+}\pi^-)$ recombine to a $D^{o}$. This is
described by
$A^{(6)}_{\rm loop}$ and is proportional to $\alpha
g^2_{D^*D\pi}\mu^{(+)}$, where $\mu^{(+)}$ is
the strength of the $D^{*+}\rightarrow D^+\gamma$ transition.

Calculating from (11) the decay width, one obtains [6] an expression
containing 13 terms, which
depend on various products $g^\alpha\mu^\beta\mu^{(+)^\gamma}$, where
$\alpha, \beta$
 have values 0 to 4 and $\gamma$
has values 0-2. At this point, the crucial step is to use the existing
experimental
information on the relative branching ratios $\Gamma(D^{*o}\rightarrow
D^o\pi^o):
\Gamma(D^{*o}\rightarrow D^o\gamma)= (61.9\pm2.9)\%: (38.1\pm 2.9)\%$ and 
$\Gamma(D^{*+}\rightarrow D^o\pi^+):
\Gamma(D^{*+}\rightarrow D^+\gamma)=(67.6\pm 0.9)\%:(1.7\pm0.6)\%$ [16,17].
This
allows us to establish $\mu\simeq 6.6g/M_{D^*}$ and $\mu_+\simeq 1.7
g/M_{D^*}$.
As a result, we were able to express $\Gamma(D^{*o}\rightarrow
D^o\gamma\gamma)$ [6] 
{\em as a function of g only}:
\begin{eqnarray}
&&\Gamma(D^{*o}\rightarrow D^o\gamma\gamma)=
[2.52\times 10^{-11} g^2 + 5.66 \times 10^{-11}g^3 \nonumber \\
&& ~~~~ + 4.76 \times 10^{-9}g^4 + 3.64 \times 10^{-10} g^5 + 1.53 \times
10^{-9} g^6] 
\ {\rm GeV} \ .%12
\end{eqnarray}
In obtaining (12) we assumed that the coupling constants are relatively
positive, as
indicated by theoretical analysis [14]. However, if we assume opposite sign
for various
pairs of couplings, we found that the changes are rather small, the reason
being
that the main contribution is given by quadratic terms.

Let us define now the branching ratio for this decay 
\begin{eqnarray}
BR(D^{*o}\rightarrow D^o\gamma\gamma) = 
\frac{\Gamma(D^{*o}\rightarrow D^o\gamma\gamma)}{\Gamma(D^{*o}\rightarrow
D^o\gamma)
+\Gamma(D^{*o}\rightarrow D^o\pi^o)}  \ .  %13
\end{eqnarray}
In the denominator, we can use for $\Gamma(D^{*o}\rightarrow D^o\pi^o)$ the
expression 
obtainable from Eq.~(4)
\begin{eqnarray}
\Gamma(D^{*o}\rightarrow D^o\pi^o) = \frac{1}{12\pi} \frac{g^2}{f^2} |
\vec{p}_\pi|^3 =
1.25 \times 10^{-4}g^2 \ {\rm GeV} %14
\end{eqnarray}
while for $\Gamma(D^{*o}\rightarrow D^o\gamma)$ we use the experimental
fact [16]
$\Gamma(D^{*o}\rightarrow D^o\pi^o):\Gamma(D^{*o}\rightarrow D^o\gamma)=
61.9:38.1$, to rexpress $\Gamma(D^{*o}\rightarrow D^o\gamma)$ in terms of 
$g^2$ as well. Thus one obtains [6]
\begin{eqnarray}
\Gamma(D^o\rightarrow \ {\rm all}) = (2.02\pm 0.12)\times 10^{-4}g^2 \ {\rm
GeV} %14
\end{eqnarray}
and as a consequence
\begin{eqnarray}
BR(D^{*o}\rightarrow D^o\gamma\gamma) =
\frac{(0.025 + 0.057g + 4.76g^2 + 0.36g^3 + 1.53 g^4)\times
10^{-9}g^2}{2.02 \times 10^{-4}g^2}
%15
\end{eqnarray}
Obviously, a measurement of this ratio will constitute a measurement of $g$.

Turning to the $B^{*o} \rightarrow B^o\gamma\gamma$ decay, one has a rather
different
situation. Firstly, the branching ratio is now defined as
\begin{eqnarray}
BR(B^{*o}\rightarrow B^o\gamma\gamma) =
\frac{\Gamma(B^{*o}\rightarrow B^o\gamma\gamma)}{\Gamma(B^{*o}\rightarrow
B^o\gamma)} %16
\end{eqnarray}
since there is no strong decay of $B^{*o}$. The quantity in Eq.~(17) turns
out to be
a function of $g$, $\mu$ and $\mu^+$ and an analysis similar to the one we
performed for $D^{*o}$ decay is not possible. Still, a certain amount of
information
is obtainable by a judicious analysis in the above parameter space [6].
However, we shall
not address this topic here.\\

\begin{center}
{\bf 4. Discussion and Summary}
\end{center}

Firstly, some remarks about the framework of the calculation of [6]. The
analysis of 
$D^{*o}\rightarrow D^o\gamma\gamma$ has been performed to leading order in
chiral perturbation
theory and mostly to leading order in a $1/M$ expansion. Corrections to the
leading order of
(2) have been studied extensively in recent years [4,14,28]. A
comprehensive treatment of such 
corrections is beyond the scope of the calculation presented in Ref.~[6].
However, one notes
that several features belonging to the next order are included in their [6]
treatment, like
the use of physical masses for the degenerate doublet of heavy mesons in
the decay
calculations and in propagators. For the latter, as well as for vertices
and normalizations the
convention of Ref.~[4] is used. Thus, the propagator of heavy vector mesons
is given by
$-i(g^{\mu\nu}- v^{\mu} v^\nu)/2[(v\cdot k) - \Delta/4]$ and of
pseudoscalar mesons by
$i/2[(v.k) + 3\Delta/4]$,
where $\Delta = M_{D^*}(M_{B^*}) - M_D(M_B)$ and $v, k$ are the velocity
and the residual 
momentum.

Additional technical points, which should be mentioned are: (i) the loop
calculations
for $D^{*o}\rightarrow D^o\gamma\gamma$ include also contributions from
intermediate 
states containing $K-$mesons, like $D^{*o}\rightarrow
(K^-D_s^{*+})\rightarrow D^o\gamma\gamma$
with the photons emitted from the virtual $K^-$'s; (ii) contributions from
diagrams containing
three heavy meson propagators were neglected, as these are very small
indeed; diagrams with two
heavy meson propagators were included; however their contribution is quite
small; (iii) the
contribution of the $\eta^o$-anomaly has been estimated and found to be
small; (iv) the off-the-mass-shell
$q^2$-dependence of the anomaly has been neglected.

The calculation described here was performed for neutral $D^{*o}$ decay
(likewise for $B^*$).
Obviously there are also the $D^{*+}\rightarrow D^+\gamma\gamma$, 
$B^{*+}\rightarrow B^+\gamma\gamma$
and $D^{*+}_s \rightarrow D_s^+\gamma\gamma$ decays. For these decays, one
has to consider
also the bremsstrahlung radiation emitted by the initial or final charged
particles. We have
estimated these decays and we found that the bremsstrahlung part is orders
of magnitude larger than
the direct one; hence a different type of analysis is required [29] and we
do not address
this here.

To summarize, we have shown [6] that the measurement of the
$D^{*o}\rightarrow D^o\gamma\gamma$ branching 
ratio constitutes an ideal tool for obtaining the magnitude of the strong
$g$-coupling,
since it can be expressed in terms of $g$ only (Eq.~16). This has been
achieved by combining
a theoretical calculation of $D^{*o}\rightarrow D^o\gamma$ using the 
Heavy Meson Chiral Lagrangian with
the experimental information relating the electromagnetic $(D^*\rightarrow
D\gamma)$
and strong $(D^*\rightarrow D\pi)$ partial decay channels. The various
theoretical estimates
put $g$ in the range $0.25 < g < 1$. Hence, from Eq.~(16) one obtains\\

\begin{center}
\begin{tabular}{cc}
$g$ & $Br[(D^{*o}\rightarrow D^o\gamma\gamma)/D^{*o}\rightarrow \ {\rm
all}]$ \\ \hline
0.25 & $1.7\times 10^{-6}$ \\
0.38 & $3.9\times 10^{-6}$ \\
0.5 & $6.9\times 10^{-6}$ \\
0.7 & $1.4\times 10^{-5}$ \\
1 & $3.3\times 10^{-5}$ \\ \\
\end{tabular}
\end{center}

These figures indicate that the suggested measurement is indeed feasible in
the not too
distant future.

\begin{center}
{\bf References}
\end{center}
\begin{enumerate}
\item M.B. Wise, {\it Phys.\ Rev.} D {\bf 45}, R2188 (1992).
\item G. Burdmann and J. Donoghue, {\it Phys.\ Lett.} B {\bf 280}, 287 (1992).
\item T.M. Yan et al., {\it Phys.\ Rev.} D {\bf 46}, 1148 (1992); 
{\bf 55}, 5851(E) (1997).
\item R. Casalbuoni et al., {\it Phys.\ Rept.} {\bf 281}, 145 (1997).
\item ACCMOR Collab., S. Barlag et al., {\em Phys.\ Lett.} B {\bf 278}, 480
(1992).
\item D. Guetta and P. Singer, hep-ph/9904454, {\em  Phys.\ Rev.} D (to be
published).
\item CUSB Collab., K Han et al., {\it Phys.\ Rev.\ Lett.} {\bf 55}, 36
(1985);
CUSB-II Collab., J. Lee-Franzini et al., {\em Phys.\ Rev.\ Lett.} {\bf 65},
2947 (1990).
\item ALEPH Collab., {\em Z.\ Phys.} C{\bf 69}, 393 (1996); DELPHI Collab.,
{\em Z.\ Phys.}
C {\bf 68}, 353 (1995); L3 Collab., {\it Phys.\ Lett.} B {\bf 345}, 589
(1995); OPAL Collab.,
{\it Z. Phys.} C{\bf 74}, 413 (1997).
\item E. Eichten et al., {\em Phys.\ Rev.} D {\bf 21}, 203 (1980); 
S. Godfrey and N. Isgur, {\em Phys.\ Rev.}
D {\bf 32}, 189 (1985); M.A. Ivanov and Yu.\ M.\ Valit, {\em Z.\ Phys.}
C{\bf 67}, 633 (1995); 
W. Jaus, {\it Phys.\ Rev.} D {\bf 53}, 1349 (1996).
\item P. Singer and G.A. Miller, {\em Phys.\ Rev.} D{\bf 34}, 825 (1989).
\item P. Cho and H. Georgi, {\em Phys.\ Lett.} B {\bf 296}, 408 (1992);
J.F. Amundson et al., {\em Phys.\ Lett.} B {\bf 296}, 415 (1992);
H.-Y. Cheng et al., {\em Phys.\ Rev.} D {\bf 47}, 1030 (1993);
P. Colangelo, F. DeFazio and G. Nardulli, {\em Phys.\ Lett.} B {\bf 316},
555 (1993).
\item N. Barik and P.C. Dash, {\em Phys.\ Rev.} D {\bf 49}, 299 (1994);
P. Colangelo, F. DeFazio and G. Nardulli, {\em Phys.\ Lett.} B {\bf 334},
175 (1994).
\item H.G. Dosch and S. Narison, {\em Phys.\ Lett.} B {\bf 368}, 163 (1996);
T.M. Aliev et al., {\em Phys.\ Rev.} D {\bf 54}, 857 (1996).
\item I. Stewart, {\em Nucl.\ Phys.} B {\bf 529}, 62 (1998).
\item I. Peruzzi et al., {\it Phys.\ Rev.\ Lett.} {\bf 37}, 569 (1976); 
Mark I Collab., G. Goldhaber et al., {\em Phys.\ Lett.} B {\bf 69}, 503
(1977);
G.J. Feldman et al., {\em Phys.\ Rev.\ Lett.} {\bf 38}, 1313 (1977).
\item Particle Data Group, C. Caso et al., {\em Europ. Phys.\ Journal}
C{\bf 3}, 1
(1998).
\item CLEO Collab., J. Bartelt et al., {\it Phys.\ Rev.\ Lett.} {\bf 80},
3919 (1998).
\item T.N. Pham, {\it Phys.\ Rev.} D {\bf 25}, 2955 (1982);
R.L. Thews and A.N. Kamal, {\it Phys.\ Rev.} D {\bf 32}, 810 (1985);
P.J. O'Donnel and Q.P. Xu, {\it Phys.\ Lett.} B {\bf 336}, 113 (1994).
\item G.A. Miller and P. Singer, {\em Phys.\ Rev.} D {\bf 37}, 2564 (1998).
\item V.M. Belyaev, V.M. Braun, A. Khodjamirian and R. Ruckl, {\em Phys.\
Rev.} D {\bf 51}, 6177
(1995).
\item N. Isgur and M.B. Wise, {\em Phys.\ Rev.} D {\bf 41}, 151 (1990);
C.A. Dominguez and N. Paver, {\em Z.\ Phys.} C{\bf 41}, 217 (1988).
\item P. Colangelo, F. De Fazio and G. Nardulli, {\it Phys.\ Lett.} B {\bf
334}, 175 (1994);
P. Colangelo, G. Nardulli and M. Pietroni, {\it Phys.\ Rev.} D{\bf 43},
3002 (1991).
\item A. Deandrea et al., {\it Phys.\ Rev.} D{\bf 58}, 034004 (1998).
\item D. Becirevic and A. Le Yaouanc, {\it JHEP} 9903:021 (1999).
\item P. Colangelo et al., {\it Phys.\ Lett.} B {\bf 339}, 151 (1994);
H.G. Dosch and S. Narison, {\it Phys.\ Lett.} B {\bf 368}, 163 (1996);
P. Colangelo and F. De Fazio, {\it Eur.\ Phys.\ J.} C {\bf 4}, 503 (1998);
A. Khodjamirian, R. Ruckl, S. Weinzierl and O. Yakovlev, {\em Phys.\ Lett.} 
B{\bf 457}, 245 (1999).
\item UKQCD Collaboration (G.M. de Divitiis et al.), {\it JHEP} 9810:010
(1998).
\item A.K. Leibovich, A.V. Manohar and M.B. Wise, {\it Phys.\ Lett.} B {\bf
358}, 347 (1995); 
{\bf 376}, 332(E) (1996).
\item C.G. Boyd and B. Grinstein, {\it Nucl.\ Phys.} B {\bf 442},. 205 (1995).
\item D. Guetta and P. Singer (to be published)
\end{enumerate}

\end{document}